# Knock Out Slow Extraction Using Betatron Sidebands at High Harmonics

Philipp Niedermayer* and Rahul Singh
*GSI Helmholtzzentrum für Schwerionenforschung, Darmstadt, Germany*

Eike Feldmeier, Christian Schömers, and Marcel Hun
*Heidelberg Ion-Beam Therapy Center (HIT), Heidelberg, Germany*

Radio frequency knock out resonant slow extraction is a standard method for extracting stored particle beams from synchrotrons by transverse excitation. Excitation signals comprising many betatron sidebands have shown to reduce intensity fluctuations of the extracted beam spill and are used at several facilities. In this contribution, the effect of individual sidebands at different harmonics on the spill quality is systematically studied using efficient noise excitation signals fully covering the respective sidebands. Particle tracking simulations show a clear correlation between the chosen excitation frequency and the spill quality. This relation is attributed to the spectrum of particle motion shortly before their extraction, and an empiric relation between the spectrum, the excitation frequency, and the resulting spill quality is given. Experiments show an additional reduction of pile-up when using higher excitation frequencies. These insights have implications for an optimal design of excitation waveforms and hardware for Knock Out extraction systems.

## I. INTRODUCTION

Radio frequency Knock Out (rf-KO) resonant slow extraction is used to extract stored particle beams from synchrotron rings [1,2]. It utilizes a working point near a sextupole-driven third-integer resonance to create a transverse, betatron amplitude dependent instability [3]. The beam is driven into the instability in a controlled manner by increasing the particle amplitude through transverse radio frequency (rf) excitation. The excitation system consists of a signal generator, amplifiers and a stripline kicker [4], inside which the electromagnetic rf field deflects traversing particles on each turn. Over several hundred turns, these transverse kicks lead to a net change of the particle amplitudes, bringing them closer to the separatrix. On crossing the separatrix, the motion of particles becomes unbound and septa are used to deflect these particles into an extraction beam line. A *spill* of extracted particles is then delivered to experiments or used for medical therapy.

A particular challenge in resonant slow extraction is posed by fluctuations of the spill intensity on various timescales. As large fluctuations reduce the efficiency of beam usage, this is referred to as a low spill quality [5,6]. Numerous efforts have been made in the past seven decades to address this particular challenge [7–16]. More recently, the simultaneous excitation of multiple betatron sidebands at different revolution harmonics has been investigated. A study at the Wakasa Wan Energy Research Center (WERC) used simulations and experiments to show that excitation signals covering the first $n$ sidebands progressively improve the spill quality as more sidebands are added [17]. These excitation signals consisted of a series of noise bands around each betatron sideband and are referred to as *many-band* signals in the following. In a study at the Heidelberg Ion Beam Therapy Center (HIT), excitation signals comprising three frequencies at two different sidebands were found to significantly improve the spill quality compared to the previously used single-band signal covering the first betatron sideband only [18]. This excitation scheme used narrowband random binary phase-shift keying (RBPSK) signals with frequencies near $1/3$ and $4/3$ of the revolution frequency.

In this contribution, the excitation of individual sidebands is systematically investigated using simulations and experiments. The effect of each sideband on the spill quality is characterized and compared to the combined excitation of two or more sidebands at low and high harmonics. Finally, the implications of these findings for the design of rf-KO excitation systems are discussed.

* Contact author: p.niedermayer@gsi.de

### A. Betatron sidebands

The betatron tune $Q_\mathrm{x} = Q_\mathrm{x,ref} + \xi_\mathrm{x}\delta$ describes the number of betatron oscillations a particle performs each turn. It is given by the working point $Q_\mathrm{x,ref}$ and the individual particle's momentum deviation $\delta = \Delta p / p_\mathrm{ref}$ via the chromaticity $\xi_\mathrm{x} = \mathrm{d}Q_\mathrm{x} / \mathrm{d}\delta$. In the following, the lowercase $q_\mathrm{x} = Q_\mathrm{x} - \lfloor Q_\mathrm{x} \rfloor < 1$ is used to denote the fractional part of the tune. At a fixed location in the synchrotron, the betatron motion modulates the transverse particle position

$$x(s,n) = \sqrt{\epsilon_\mathrm{x}\beta_\mathrm{x}(s)}\cos(2\pi Q_\mathrm{x} n + \mu_\mathrm{x}(s) + \Phi_\mathrm{x})$$

where $n \in \mathbb{N}_0$ is the integer turn number and $s < C$ the location along the orbit with circumference $C$. The modulation amplitude is given by $\epsilon_\mathrm{x}$ and the beta-function $\beta_\mathrm{x}(s)$. The phase is given by $\Phi_\mathrm{x}$, $Q_\mathrm{x}$ and the phase advance function $\mu_\mathrm{x}(s)$ with $\mu_\mathrm{x}(C) = 2\pi Q_\mathrm{x}$. Due to this turn-by-turn modulation, the spectrum of the transverse beam motion is dominated by the well known betatron sidebands at

$$f_\pm = (h \pm k q_\mathrm{x}) f_\mathrm{rev} \tag{1}$$

where $h \in \mathbb{N}_0$ is the nearby harmonic of the revolution frequency $f_\mathrm{rev}$ and $k \in \mathbb{N}$ accounts for the appearance of second and higher order betatron sidebands. These higher order sidebands are the consequence of an additional frequency modulation caused by the nonlinear motion in the vicinity of the sextupole-driven third-integer resonance [19]. In the following, the discussion is limited to the dominant first-order sidebands for which $k = 1$.

### B. Spill quality

To assess the quality of a spill, the intensity of the extracted beam is recorded as the particle count $N$ per time interval $\Delta t_\mathrm{count}$. Fluctuations of the intensity are then evaluated over a time period $\Delta t_\mathrm{eval} \gg \Delta t_\mathrm{count}$ using the mean $\mu_\mathrm{N} = \langle N \rangle$ and standard deviation $\sigma_\mathrm{N} = \sqrt{\langle (N - \langle N \rangle)^2 \rangle}$ of counts. From these, the coefficient of variation $c_\mathrm{v}$ and the equivalent spill duty factor $F$ are calculated as [6]

$$c_\mathrm{v} = \frac{\sigma_\mathrm{N}}{\mu_\mathrm{N}} = \sqrt{\frac{\langle N^2 \rangle}{\langle N \rangle^2} - 1} \qquad F = \frac{1}{1 + c_\mathrm{v}^2} = \frac{\langle N \rangle^2}{\langle N^2 \rangle}$$

where a smaller $c_\mathrm{v}$ and larger $F$ correspond to a better spill quality. For the ideal case of a Poisson process with $\sigma_\mathrm{N} = \sqrt{\mu_\mathrm{N}}$, the coefficient of variation is $c_\mathrm{v} = \langle N \rangle^{-1/2}$. In the following, $\Delta t_\mathrm{eval} = 1000 \Delta t_\mathrm{count} = 100$ ms is used.

In addition, the delay between the extraction of consecutive particles is of interest to quantify the amount of pile-up. Considering a detector system which requires a separation time of at least $\Delta t_\text{pile-up}$ to distinguish and resolve single events, pile-up is defined by the fraction of inter-particle-delays smaller than $\Delta t_\text{pile-up}$. In the following, $\Delta t_\text{pile-up} = 20$ ns is used, motivated by the pulse width observed in the experimental data.

### C. Simulation framework

Particle tracking simulations of the rf-KO extraction process are performed with Xsuite [20], using the lattice of the HIT synchrotron [21] with the parameters listed in table I. Thereby, an unbunched (coasting) beam comprising $10^6$ particles evenly distributed along the circumference of the synchrotron is tracked.

The individual tune of particles deviates from the working point $Q_\mathrm{x,ref}$ due to chromatic and amplitude detuning [22], and is limited by the nearby third-integer resonance $Q_\mathrm{res} = 5/3$ where particles are extracted. To cover this range, an excitation signal with a rectangular spectral density with full bandwidth $\Delta Q_\mathrm{ex} = 0.02$ is used. A band-filtered noise signal is chosen because of its homogeneous spectral density. This avoids the large influence of the central frequency on the spill quality as observed for narrowband excitation signals [23], which would dominate and hinder the investigation of the behaviour at different sidebands. The amplitude of the excitation signal is controlled with a feedback system to achieve a constant spill rate of $\dot{N} = 5 \cdot 10^5$ particles/s as given by the number of ions and the extraction time of 2 s used in the simulations. To model disturbances from the electrical grid, ripples on the quadrupole magnets comprising the dominant harmonics of the mains frequency at 100 Hz, 250 Hz and 300 Hz as well as noise up to 10 kHz with a relative magnitude of $10^{-5}$ are added [24,25].

TABLE I. Beam, optics and excitation parameters

| | | |
|---|---|---|
| **Ion** | Carbon $^{12}C^{6+}$ | |
| **Energy** | $E_\mathrm{kin}/m = 251$ MeV/nucleon | |
| **Momentum** | $p_\mathrm{ref} = 8.74$ GeV/c | |
| **Rigidity** | $B\rho = 4.86$ T m | $E\rho = 897$ MV |
| **Lorentz factor** | $\gamma = 1.27$ | $\beta = 0.616$ |
| **Rev. frequency** | $f_\mathrm{rev} = 2.842\,719$ MHz | |
| **Norm. emittance** | $\varepsilon_\mathrm{xn} = 1.5$ mm mrad | |
| **Momentum spread** | $\delta_{1\sigma} = 1 \cdot 10^{-3}$ | |
| **Betatron tune** | $Q_\mathrm{x,ref} = 1.6790$ | $Q_\mathrm{y,ref} = 1.7550$ |
| **Chromaticity** | $\xi_\mathrm{x} = -1.59$ | $\xi_\mathrm{y} = -1.99$ |
| **Norm. sextupole** | $S = 27.9\ \mathrm{m}^{-1/2}$ | $\mu_\mathrm{S} = -2.9°$ |
| **Transition** | $\gamma_\mathrm{tr} = 1.82$ | $\alpha_p = 0.30$ |
| **Slip factor** | $\eta = \gamma^{-2} - \gamma_\mathrm{tr}^{-2} = 0.32$ | |
| **Excitation signal** | Band-filtered noise | |
| **Central frequency** | $Q_\mathrm{ex} = f_\mathrm{ex}/f_\mathrm{rev} = h \pm 0.327$ | |
| **Bandwidth** | $\Delta Q_\mathrm{ex} = \Delta f_\mathrm{ex}/f_\mathrm{rev} = 0.02$ | |

## II. SPILL QUALITY FOR SINGLE-BAND EXCITATION

Figure 1 shows the spill quality as a function of the excited sideband obtained from the simulation of the rf-KO extraction process. A significant improvement is observed when exciting sidebands of higher revolution harmonics. This trend is in agreement with the findings obtained for many-band excitation at WERC [17], however, the improvement is already observed for single-band excitation at the higher frequencies. The excitation power required at these frequencies is only marginally larger compared to lower frequencies (about 3 dB, see Fig. 1, bottom). At the commonly used lower excitation frequencies near the first few harmonics, a systematic difference between the excitation of upper and lower sidebands is observed. In the case presented here, the spill quality is improved by exciting a sideband at $h - q_x$ instead of a nearby sideband at $h + q_x$. With about a factor two, this effect is largest for the two sidebands around $h = 2$. However, as shown in section II B, the role of the sidebands, the magnitude and location of the maximum value of $c_v$ and the limit of $c_v$ towards high excitation frequencies are specific to the properties of the beam and lattice.

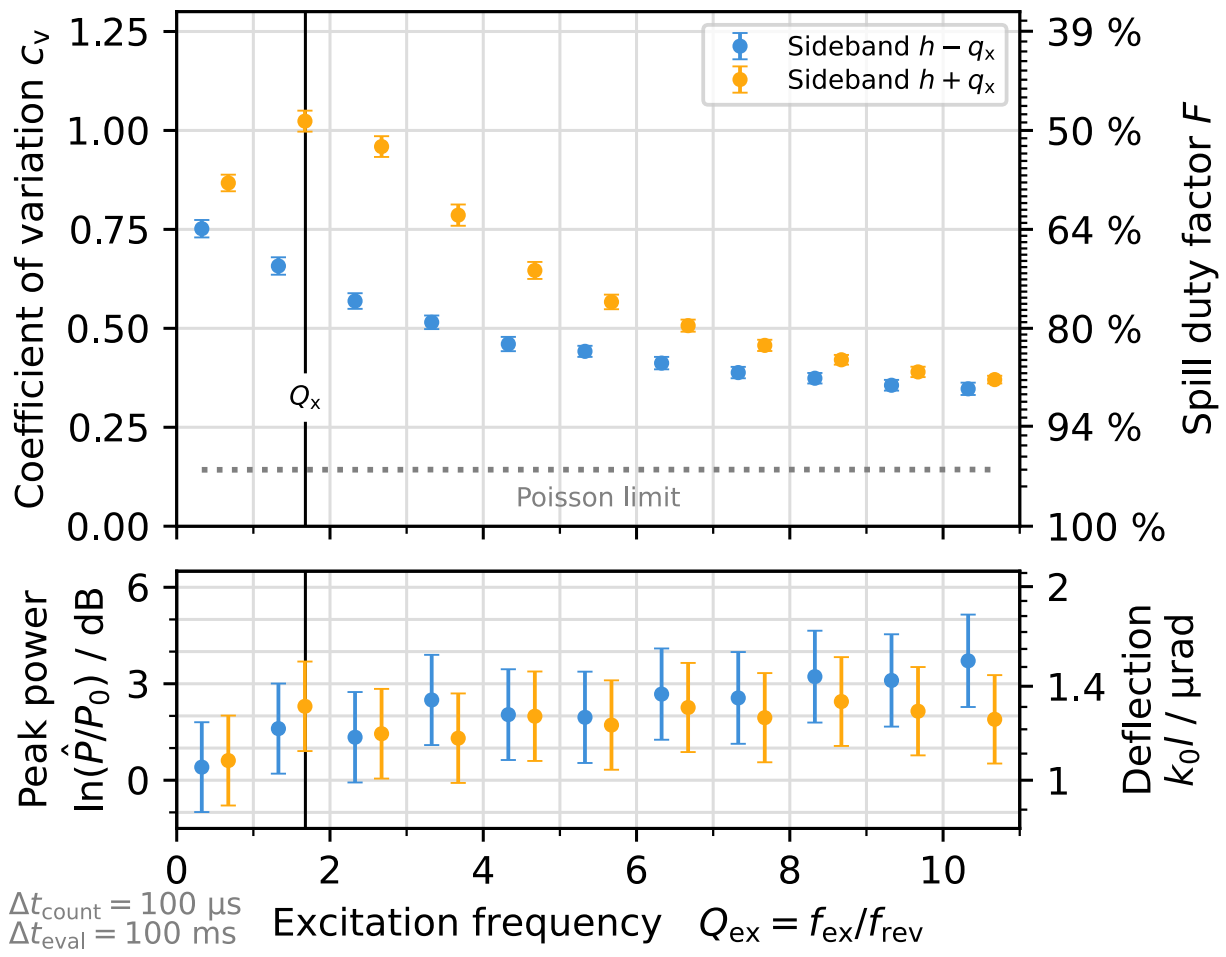

FIG. 1: Spill quality (top) and excitation power (bottom) as function of excitation frequency. Simulation of single-band signals.

### A. Imprint of excitation frequency onto spill

In resonant slow extraction, the spill quality is dominated by intensity fluctuations of typically up to a few 10 kHz. In addition, for rf-KO, the excitation frequency and its harmonics are imprinted onto the spill to some extent. This is an intrinsic property of the rf-KO method and the consequence of coherent beam oscillations at these frequencies induced by the excitation. The resulting spill structures of typically a few MHz are not relevant for facilities like HIT dedicated to hadron cancer therapy. However, for high intensity particle physics experiments such as those at GSI and the Facility for Antiproton and Ion Research (FAIR), the imprint of these frequencies is undesired as it can facilitate the extraction of particles in short succession, leading to pile-up and inefficient beam usage. This is especially problematic when the dominant low frequency fluctuations cause a momentary increase of the extraction rate, temporarily increasing the probability for pile-up.

The simulations show that the imprint of the excitation frequency can be mitigated by exciting an appropriate sideband at a higher revolution harmonic. Figure 2 shows how both the fluctuations at low frequencies and the imprint at the excitation frequency are suppressed in the same manner as the spill quality improves. For high intensity particle physics experiments, this means that the probability for pile-up is reduced in two ways: First, the suppression of low frequency fluctuations avoids the temporary increase of the extraction rate that facilitates pile-up statistically. Second, the mitigation of the high frequency imprint additionally reduces the probability for particles being extracted in short succession. As a result, pile-up is reduced as a function of the excitation frequency in analogy to the spill quality, as can be seen by comparing Figs. 1 and 3. Note that even when the high frequency imprint of the excitation onto the spill is not mitigated, a reduction of pile-up as function of the *absolute* excitation frequency is expected when the excitation frequency exceeds the average extraction rate. This is discussed in section V B, but it can not explain the differences in pile-up observed for excitation of upper and lower sidebands at low frequencies as observed in Fig. 3. This is an effect of the reduced frequency imprint.

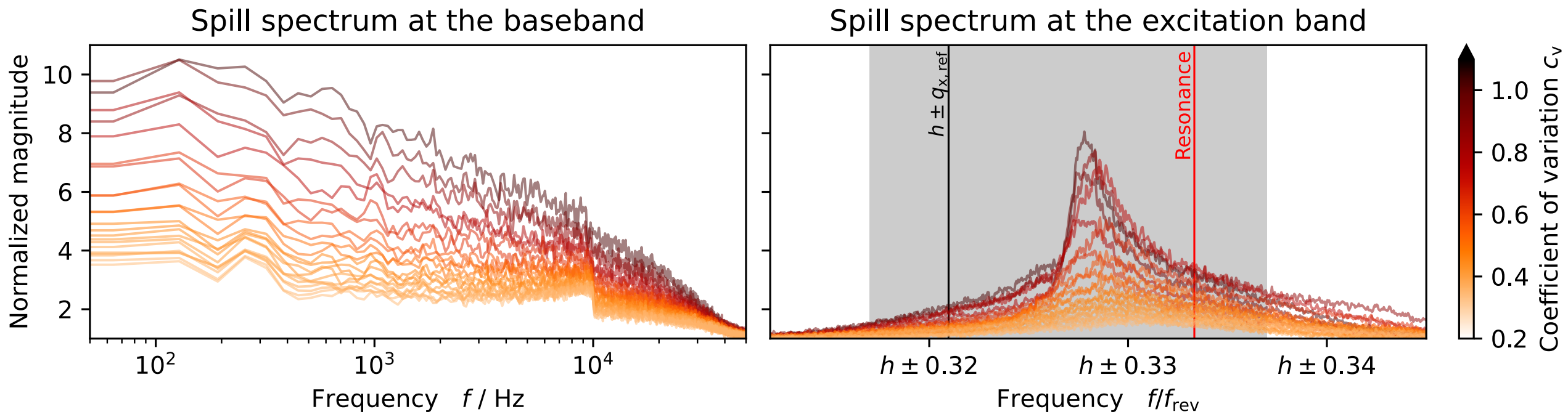

FIG. 2: Spill spectra at the baseband (left) and at the respective excitation band (right). The color correlates with the coefficient of variation according to Fig. 1, i.e. lighter shades correspond to higher excitation frequencies. Vertical lines mark the tune and resonance in the respective band; the gray shading marks the extent of the excitation band. Simulation result.

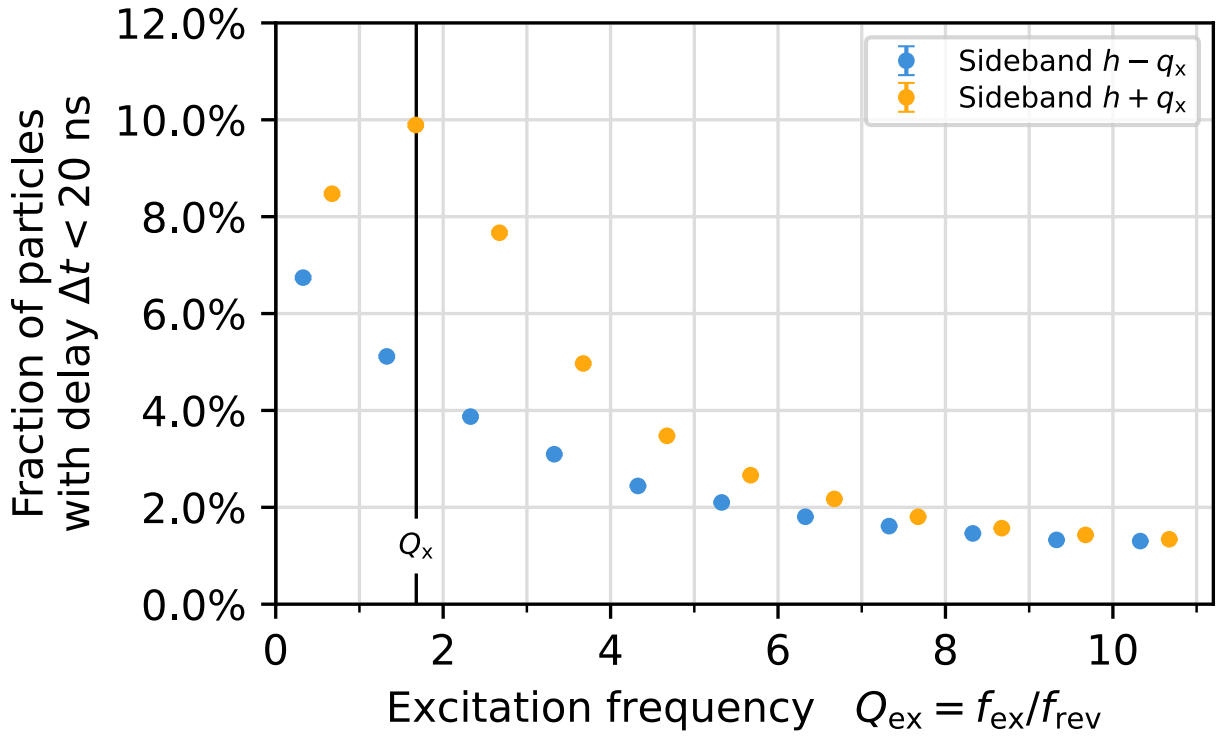

FIG. 3: Amount of pile-up as function of excitation frequency for the case of exciting a single betatron sideband. Pile-up is expressed in terms of the fraction of particles which are separated by less than $\Delta t_{\text{pile-up}} = 20$ ns. Simulation result.

### B. Influence of beam and optics parameters

To further study the role of upper and lower sidebands, it is useful to generalize the description of the excitation signal to negative frequencies. For a real-valued excitation signal, frequencies below zero are the mirror image of the positive spectrum and result in the same physical signal being applied to the beam. By changing the sign of the excitation frequency for the sidebands at $h - q_x$, the treatment of upper and lower sidebands can be unified. With the excitation frequency defined as $Q_{ex} \approx h_0 + q_x$ where $h_0 \in \mathbb{Z}$ can be negative, the improvement of the spill quality becomes symmetric with respect to the maximum, saturating towards large $|h_0|$. However, this functional dependency is subject to a number of beam and machine parameters, which is discussed in the following.

The relative **momentum spread** $\delta_{1\sigma}$ of the beam determines how rapidly the spill quality improves as sidebands further away from the maximum are excited (Fig. 4, top). In the limit of a vanishing momentum spread, no difference between the excitation of different sidebands is observed. A larger momentum spread is beneficial and reduces spill fluctuations not only at lower excitation frequencies but especially when exciting sidebands at higher frequencies. Also, the difference in spill quality between upper and lower sidebands is magnified as a function of the momentum spread. The required excitation power is independent of the momentum spread, as long as the excitation covers the momentum dependent tune spread of the beam (here $\delta_{1\sigma} \lesssim 2 \cdot 10^{-3}$).

The **chromaticity** $\xi_x$ of the lattice influences the location of the maximum value of $c_v$ (Fig. 4, middle). An inversion of the sign of the chromaticity causes the roles of upper and lower sidebands to interchange, while for zero chromaticity they behave identically and the spill quality improves purely as a function of excitation frequency or harmonic. In analogy to the momentum spread, a large chromatic tune spread exceeding the excitation band leads to an increased power requirement for $|\xi| \gtrsim 2$.

The **slip factor** $\eta = \gamma^{-2} - \gamma_{tr}^{-2}$ depends both on the beam energy $E = \gamma E_0$ and the lattice's gamma transition $\gamma_{tr}$. It influences the sharpness of the spill quality improvement by acting as a scaling factor on $Q_{ex}$ (Fig. 4, bottom). For a larger absolute value of the slip factor (smaller beam energy) the spill quality improvement has a stronger correlation with increasing excitation frequency. For small excitation frequencies at the first sideband, however, the spill quality is largely independent of the slip factor. While the excitation power naturally scales with the beam energy, its dependence on the excitation frequency is independent of the slip factor.

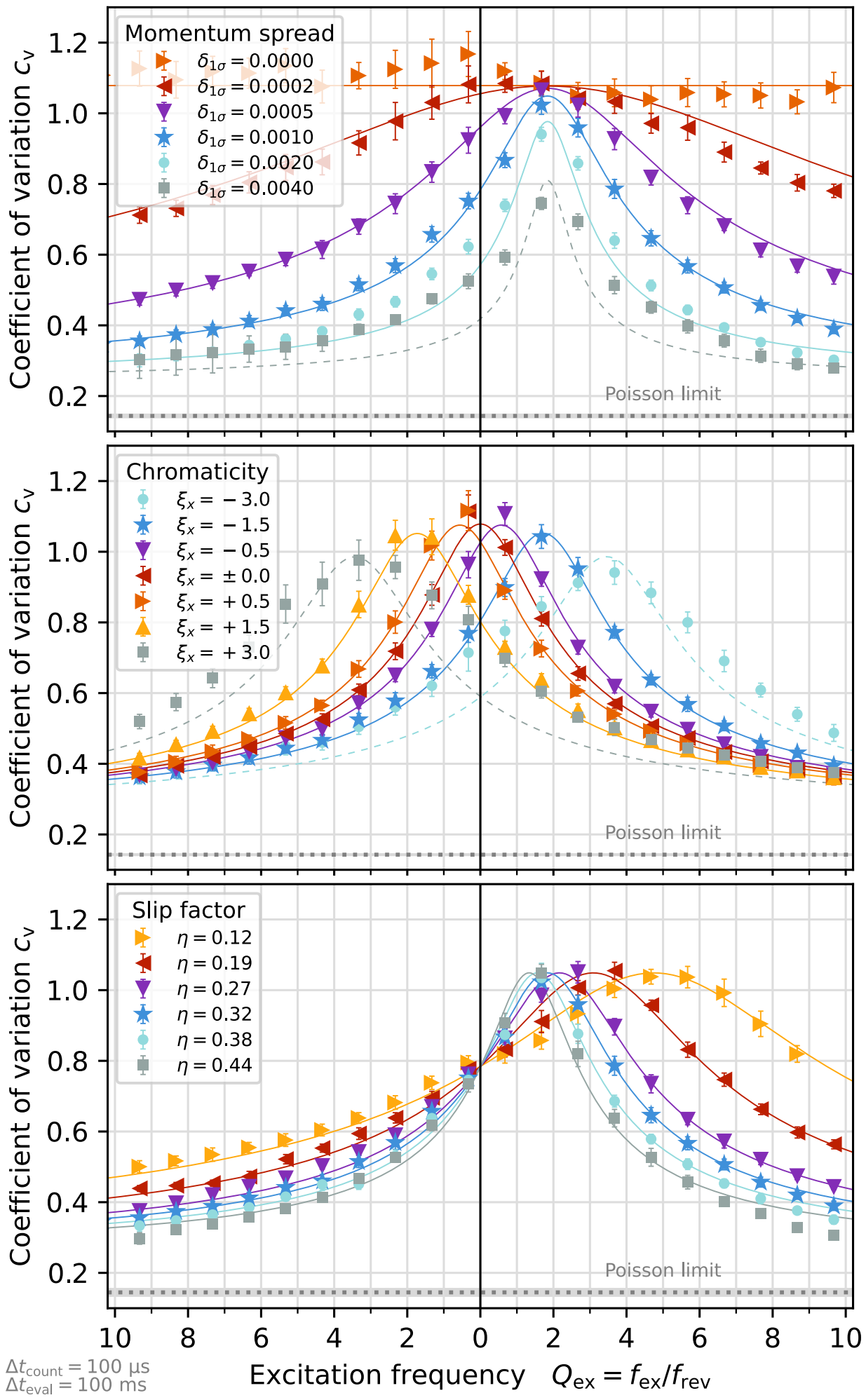

FIG. 4: Spill quality as function of excited sideband and momentum spread (top), chromaticity (middle) and slip factor (bottom). In each plot, the case of $\delta_{1\sigma} = 0.001$, $\xi = -1.5$ and $\eta = 0.32$ (blue stars) corresponds to Fig. 1. Each data point corresponds to a distinct simulation where one of these parameters is varied systematically. Thereby, the slip factors correspond to different beam energies between $E_{kin} = 150$ MeV/u for $\eta = 0.44$ and $500$ MeV/u for $\eta = 0.12$. For better visualization, sidebands at $h - q_x$ are shown as mirrored to the left half, while sidebands at $h + q_x$ are shown on the right half. The solid lines are the result of a multivariate fit using Eqs. (5) and (6). Where the tune spread exceeds the excitation band width, the line is dashed.

## III. SCHOTTKY SPECTRA OF STRONG NONLINEAR BEAM MOTION

The functional dependence of the spill quality on the excitation frequency as well as the momentum spread, tune, chromaticity and slip factor discussed in section II B suggests, that it is related to the spectral properties of the respective sidebands as observed in Schottky spectra of a corresponding non-excited beam. For the case of linear particle motion, transverse Schottky spectra are well understood [26]. Particles perform harmonic betatron oscillations and the sideband frequency follows Eq. (1) for $k = 1$. Thereby each particle contributes a momentum-dependent frequency of

$$\frac{f_\pm}{f_\mathrm{rev}} = [h \pm q_\mathrm{x} \pm \xi_\mathrm{x}\delta](1+\eta\delta) \tag{2}$$

with the parameters from table I and the plus and minus signs for the upper and lower sidebands respectively. Considering a beam with a 1-sigma momentum spread of $\delta_{1\sigma}$ and neglecting small $\delta^2$ terms, the 1-sigma bandwidth of each sideband is given by

$$\frac{\Delta f_\pm}{f_\mathrm{rev}} \approx \left[(h \pm q_\mathrm{x})\eta \pm \xi_\mathrm{x}\right]\delta_{1\sigma} \tag{3}$$

### A. Nonlinear and hollow beam spectra

To adapt the linear description of Schottky spectra to the nonlinear case, an empirical amplitude detuning term is added to Eq. (2). This term takes the amplitude dependent average detuning $\mathcal{V}_\mathrm{a}$ towards the third-integer resonance into account:

$$\frac{f_\pm}{f_\mathrm{rev}} \approx \left[h \pm q_\mathrm{x} \pm \xi_\mathrm{x}\delta \pm \mathcal{V}_\mathrm{a}\left(1-\frac{\xi_\mathrm{x}\delta}{\Delta_\mathrm{res}}\right)\right](1+\eta\delta) \tag{4}$$

where $\Delta_\mathrm{res} = Q_\mathrm{res} - Q_\mathrm{x,ref}$ is the nominal tune distance to the resonance and the factor in round brackets corrects for the fact that this tune distance is reduced for an off-momentum particle.

To estimate the sideband width of a beam in the nonlinear case not only the momentum distribution but also the distribution in $\mathcal{V}_\mathrm{a}$ must be taken into account. While this generally depends on the amplitude distribution of particles in transverse phase space, for simplicity an average value of $\langle\mathcal{V}_\mathrm{a}\rangle$ and a 1-sigma spread of $\Delta\mathcal{V}_\mathrm{a}$ is assumed in the following to estimate the resulting frequency distribution. In addition to the average amplitude detuning, the nonlinear dynamics also cause a modulation of the betatron frequency [19,27]. To account for the resulting additional frequency spread, an additional phase detuning term $\Delta\mathcal{V}_\mathrm{p}$ is added. The 1-sigma bandwidth of each sideband can thus be estimated empirically (see appendix) as

$$\frac{\Delta f_\pm}{f_\mathrm{rev}} \approx \Delta\mathcal{V}_\mathrm{p} + \sqrt{\left[\left(h \pm q_\mathrm{x} \pm \langle\mathcal{V}_\mathrm{a}\rangle\right)\eta \pm \xi_\mathrm{x}\left(1-\frac{\langle\mathcal{V}_\mathrm{a}\rangle}{\Delta_\mathrm{res}}\right)\right]^2 \delta_{1\sigma}^2 + \Delta\mathcal{V}_\mathrm{a}^2 + c_\mathrm{cor}\left(\eta - \frac{\xi_\mathrm{x}}{\Delta_\mathrm{res}}\right)^2 \delta_{1\sigma}^2\Delta\mathcal{V}_\mathrm{a}^2} \tag{5}$$

where $\langle\mathcal{V}_\mathrm{a}\rangle$, $\Delta\mathcal{V}_\mathrm{a,p}$ and $c_\mathrm{cor}$ serve as fit parameters. The bilinear correction factor $c_\mathrm{cor}$ is introduced to correct for the approximation of describing $\delta$ and $\mathcal{V}_\mathrm{a}$ by independent random distributions with mean and 1-sigma spread, as in reality the chromatic detuning and amplitude detuning are coupled. Fig. 5 shows the sideband widths as obtained from simulated Schottky spectra for a full and a hollow beam in comparison to the predictions from the linear and empirically adopted nonlinear theory. For the *full beam*, the fit using Eq. (5) and the parameters from table I yields an amplitude detuning $\langle \mathcal{V}_\mathrm{a} \rangle$ and $\Delta\mathcal{V}_\mathrm{a}$ with values smaller than their uncertainty, indicating that the amplitude detuning is negligible as the spectrum is dominated by the densely populated beam core with small amplitudes. The phase detuning $\Delta\mathcal{V}_\mathrm{p} = (1.93 \pm 0.18) \cdot 10^{-3}$ is not negligible and can explain the increase of the sidebands frequency spread with respect to the linear theory in reasonable agreement considering the simplistic empiric approximation of the complex nonlinear dynamics involved. However, as the spectrum is dominated by the densely populated beam core, it can not describe the spill quality improvement in Fig. 1.

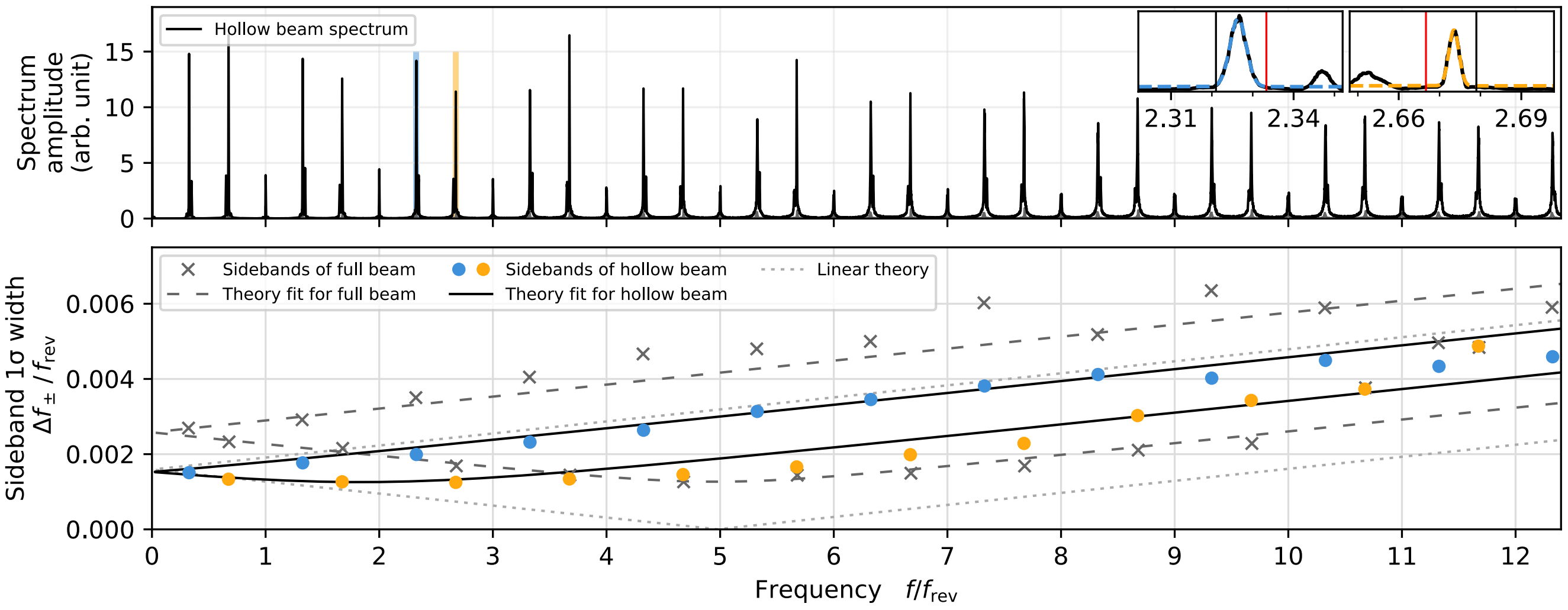

FIG. 5: Top: simulated transverse Schottky spectrum of a hollow beam with betatron sidebands. Insets show zooms for the two highlighted sidebands with Gaussian fits. The small additional peaks visible in the insets are the sidebands of second order. Bottom: 1-sigma widths of the sidebands as obtained from the Gaussian fits for a full and hollow beam scenario. The dotted gray line shows the width expected from Eq. (3) and the dashed and solid lines are fits using Eq. (5) and the parameters from table I.

The study of a *hollow beam* distribution as depicted in Fig. 6 is motivated by the fact that the extraction process is governed by particles near the separatrix performing strongly nonlinear motion. The Schottky spectrum obtained from a simulation of such a hollow beam (without rf-KO excitation) is also shown in Fig. 5. Simulations of different hollow beam distributions comprising the outer 10 %, 25 % or 44 % of particles near the separatrix (in terms of the Hamiltonian) confirm the robustness and convergence of the depicted widths (results not shown for brevity). In this case, the width of the sidebands shows the same behavior as the spill quality obtained by exciting the respective sideband (Fig. 1): The smallest width corresponds to the largest spill fluctuations near $h = 2$, and the spill fluctuations decrease as the width increases towards higher frequencies, resulting in an improved spill quality.

### B. Correlation with spill quality

The correlation between the spill quality and the width of the sidebands in the spectra of a hollow beam also remains valid if the parameters discussed in section II B are changed. Fig. 7 shows this correlation for various optics settings. The functional dependency found empirically based on the data is:

$$c_\mathrm{v} = c_{\mathrm{v},\infty} + \frac{a}{\Delta f_\pm / f_\mathrm{rev} - b} \tag{6}$$

where $c_\mathrm{v}$ is the resulting spill quality by means of the coefficient of variation, $\Delta f_\pm / f_\mathrm{rev}$ is the 1-sigma width of the sideband as obtained from hollow beam simulations without excitation, and $a$, $b$ and $c_{\mathrm{v},\infty}$ are fit parameters. The fit shown in Fig. 7 yields $a = (4.4 \pm 0.4) \cdot 10^{-4}$, $b = (7.0 \pm 0.4) \cdot 10^{-4}$ and the spill quality limit $c_{\mathrm{v},\infty} = 0.24 \pm 0.02$.

By combining Eq. (5) and (6) and performing a meta-fit, one can find the detuning parameters $\langle \mathcal{V}_\mathrm{a} \rangle$, $\Delta\mathcal{V}_\mathrm{a,p}$ and $c_\mathrm{cor}$ representative for the spectral composition of the particles governing the extraction process. The resulting dependency is shown by the solid lines in Fig. 4 and as theory fit for the hollow beam case in Fig. 5. For the meta-fit, all data points from the considered parameter space are used, except for the largest momentum spread and chromaticities as for these the beam spectrum becomes wider than the excitation bandwidth (dashed lines in Fig. 4). The meta-fit yields $\langle \mathcal{V}_\mathrm{a} \rangle = (-7.80 \pm 0.05) \cdot 10^{-3}$, $\Delta\mathcal{V}_\mathrm{a} = (4.82 \pm 0.14) \cdot 10^{-4}$, $\Delta\mathcal{V}_\mathrm{p} = (7.55 \pm 0.13) \cdot 10^{-4}$ and $c_\mathrm{cor} = (2.2 \pm 0.3) \cdot 10^{-6}$. These values are in good agreement with the expected magnitude considering the distance to the resonance of $\Delta_\mathrm{res} = -0.0123$. The average detuning of $|\langle \mathcal{V}_\mathrm{a} \rangle / \Delta_\mathrm{res}| = 63\,\%$ as obtained from the fit matches the prediction from the Kobayashi theory of 60 % or more for particles near the resonance [28].

### C. Interpretation

This correlation can be understood as follows: A larger sideband width in the hollow beam spectrum means that the frequency spread of the nonlinear betatron motion particles perform near the third-integer resonance separatrix is larger. A broadband excitation with frequencies ranging from the working point to the resonance can then couple more homogeneously to the beam as particles transit from the core to the separatrix. The result is a "resistive" excitation process with a larger net energy transfer to the beam. On the other hand, if the resonant frequency spread of the beam is small compared to the excitation bandwidth, the process is "reactive", resulting in a perturbation of the particle motion prior to extraction rather than an energy transfer. This perturbation is observed as a stronger imprint of the non-resonant part of the excitation spectrum on the spill. It can also be understood as a demodulation of the excitation spectrum by the beam, causing the "overhanging" frequencies in the excitation spectrum not matched to the beam spectrum to be transformed onto the spill spectrum in baseband as can be seen in Fig. 2 (left). While it seems reasonable to mitigate this perturbation also for smaller sidebands by using an excitation signal with a smaller bandwidth, such signals cannot efficiently drive the rf-KO extraction, because they cannot account for the intrinsic amplitude detuning from the beam core to the separatrix. As such, these methods rely on a higher excitation power, which in turn can amplify the perturbation and negate the desired effect, although the specific details for narrow-band excitation signals also depend on thir exact centre frequency. Only a sinusoidal excitation signal with zero bandwidth can achieve the desired effect, as it is not modulated in the first place and has no spectral bandwidth which could perturb the extraction process — even with the high excitation strength required. In fact, by speeding up the extraction and separatrix crossing for individual particles, the sinusoidal excitation can even suppress perturbations from power supply induced separatrix size fluctuations or those resulting from combinations with broadband excitation signals [23].

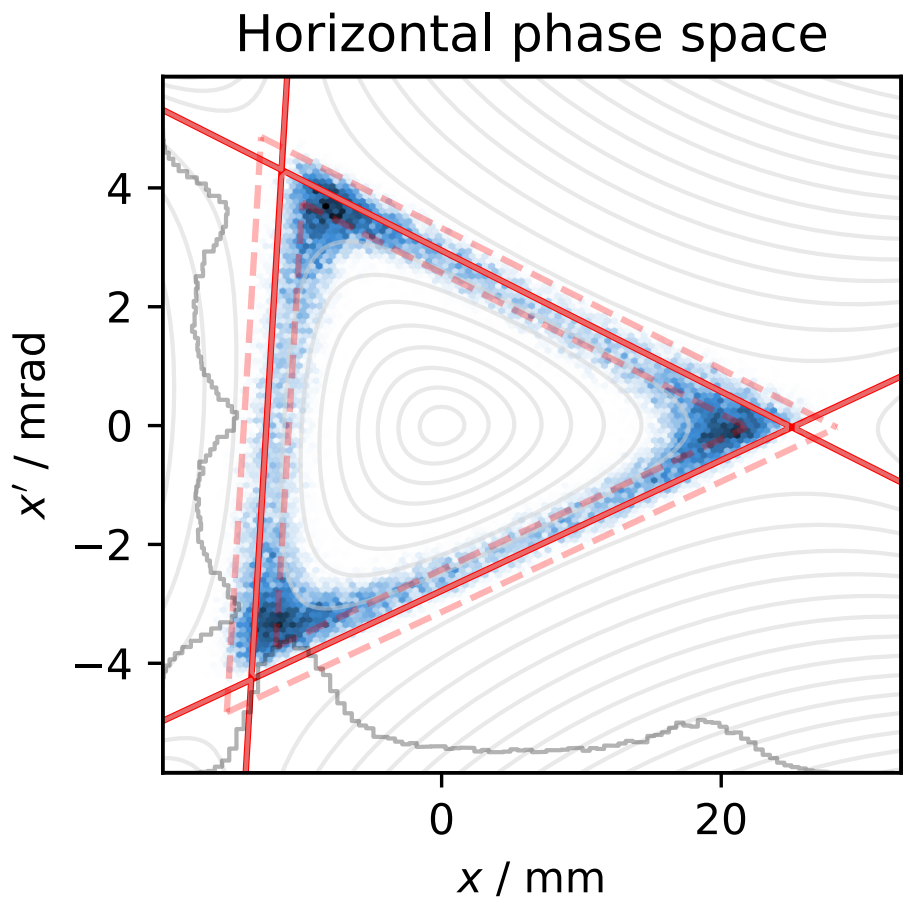

FIG. 6: Hollow beam distribution in horizontal phase space (corrected for closed orbit and dispersion) occupying the outer 10 % in terms of the Hamiltonian (equipotentials marked by thin gray lines). Red lines mark the separatrix of the third-integer resonance for momentum offsets of $\delta = 0$ (solid) and $\delta = \pm 10^{-3}$ (dashed).

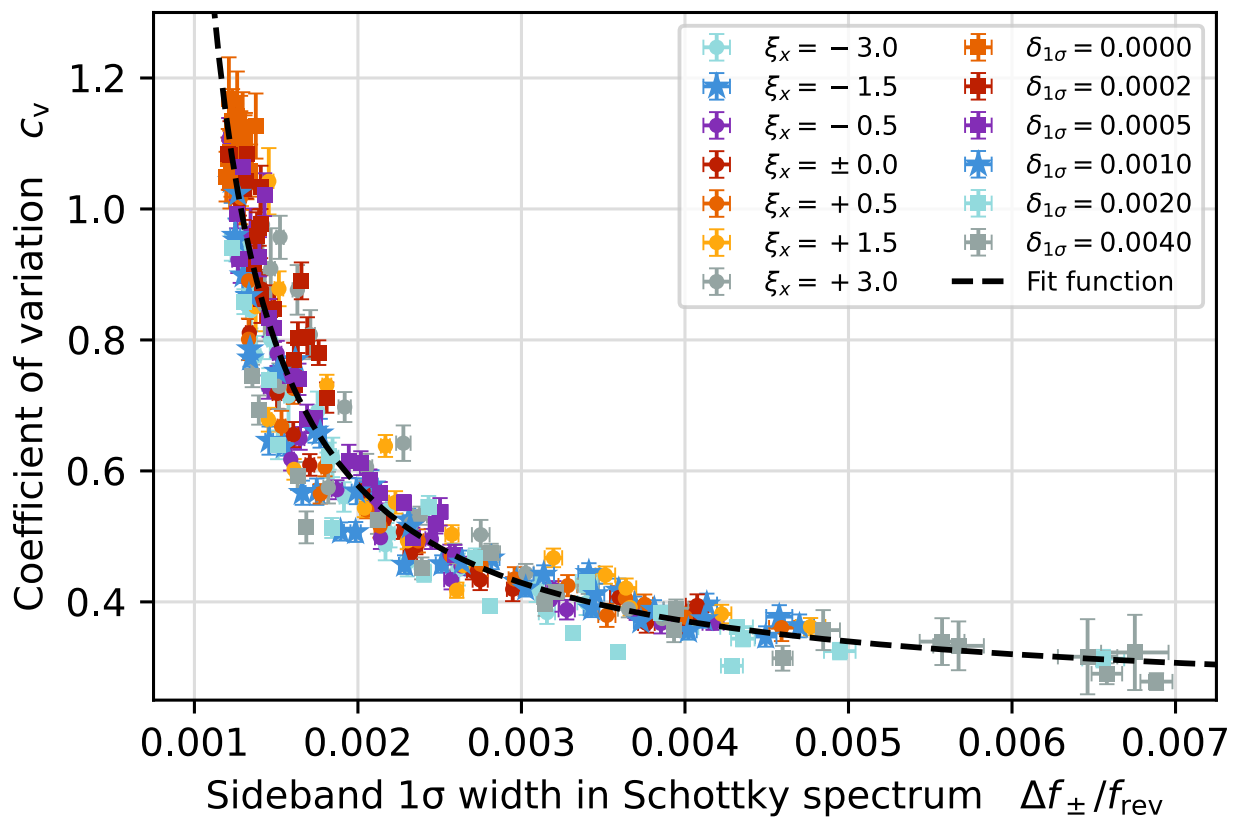

FIG. 7: Correlation between the 1-sigma width of sidebands in the Schottky spectrum of a hollow beam (compare Fig. 5) and the spill quality resulting from its excitation, obtained from simulations with various optics settings (compare Fig. 4). The excitation bandwidth is $\Delta Q_\mathrm{ex} = 0.02$. The dashed fit function is described in the text.

Generally, to maximize the net energy transfer throughout the transition of particles from the core to the separatrix, a broadband excitation signal has to be used. In this case, exciting the beam at a sideband where the spectrum of particle motion is broader has the same effect of mitigating the perturbation of the extraction process and resulting imprint of the excitation, since the excitation matches the spectrum of particles near the separatrix better, which results in a better spill quality as discussed above. For this process, the width as obtained from a hollow beam distribution is a determining factor, since only the spectrum of particles near the separatrix is relevant for the extraction process.

## IV. COMPARISON TO DUAL AND MANY-BAND EXCITATION

Figure 8 compares the results obtained for single-band excitation as discussed above to different multi-band excitation methods: Dual-band signals as the combination of two single-band signals at different sidebands; and many-band signals as the sum of $n$ single-band signals at the first $n$ sidebands. For the many-band cases, the spill quality improves with the number of bands, which is in agreement with the previous findings at HIT [18] and WERC [17]. In view of the single-band excitation study, this improvement can be attributed to the contribution of the covered single sidebands at higher harmonics. In a simplified picture, the resulting spill quality can be understood as the average obtained from each distinct excitation sideband. This is demonstrated in Fig. 9 where the average width of the respectively excited sidebands from the hollow beam Schottky spectrum is correlated with the resulting spill quality. The dependency follows the same functional relation found previously for the single-band study (Fig. 7). Since the many-band excitation also always contains contributions from the lower harmonics which have a worse spill quality, the spill quality improvement is weaker compared to the single-band excitation. Accordingly, dual-band excitation at (only) the high harmonics result in a better spill quality compared to the many-band case containing also all the lower harmonics.

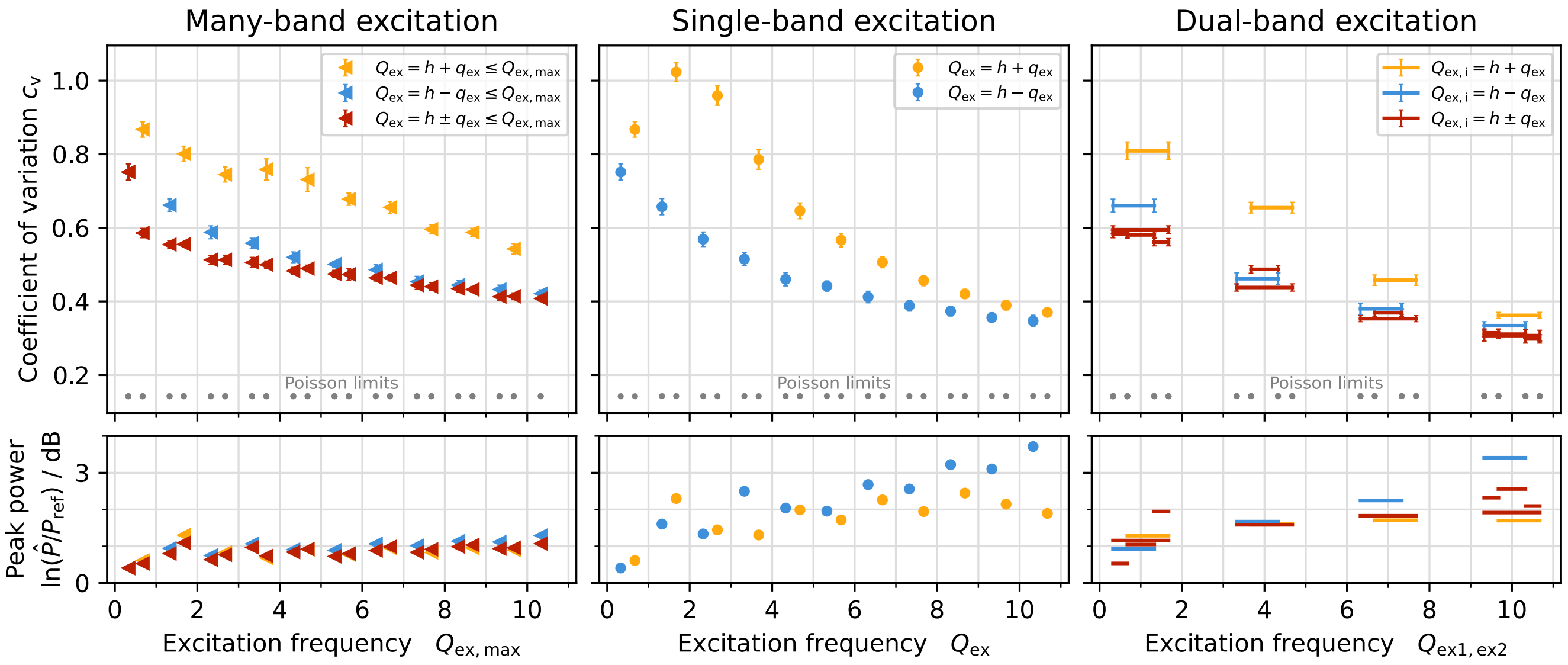

FIG. 8: Spill quality (top) and excitation power (bottom) as function of excitation frequency for excitation of sidebands in different combinations. Left: Simultaneous excitation at many sidebands, i.e. the first $n$ upper, first $n$ lower, or all first $n$ sidebands up to the indicated maximum frequency. Center: Excitation of single sidebands (same data as Fig. 1 for reference). Right: Simultaneous excitation at two sidebands, i.e. two upper, two lower, or one upper and one lower sideband respectively at the pairwise indicated frequencies. The power $P \propto (k_0 l)^2$ is shown relative to the reference deflection $k_0 l_{ref} = 1$ µrad. Simulation according to the parameters listed in table I.

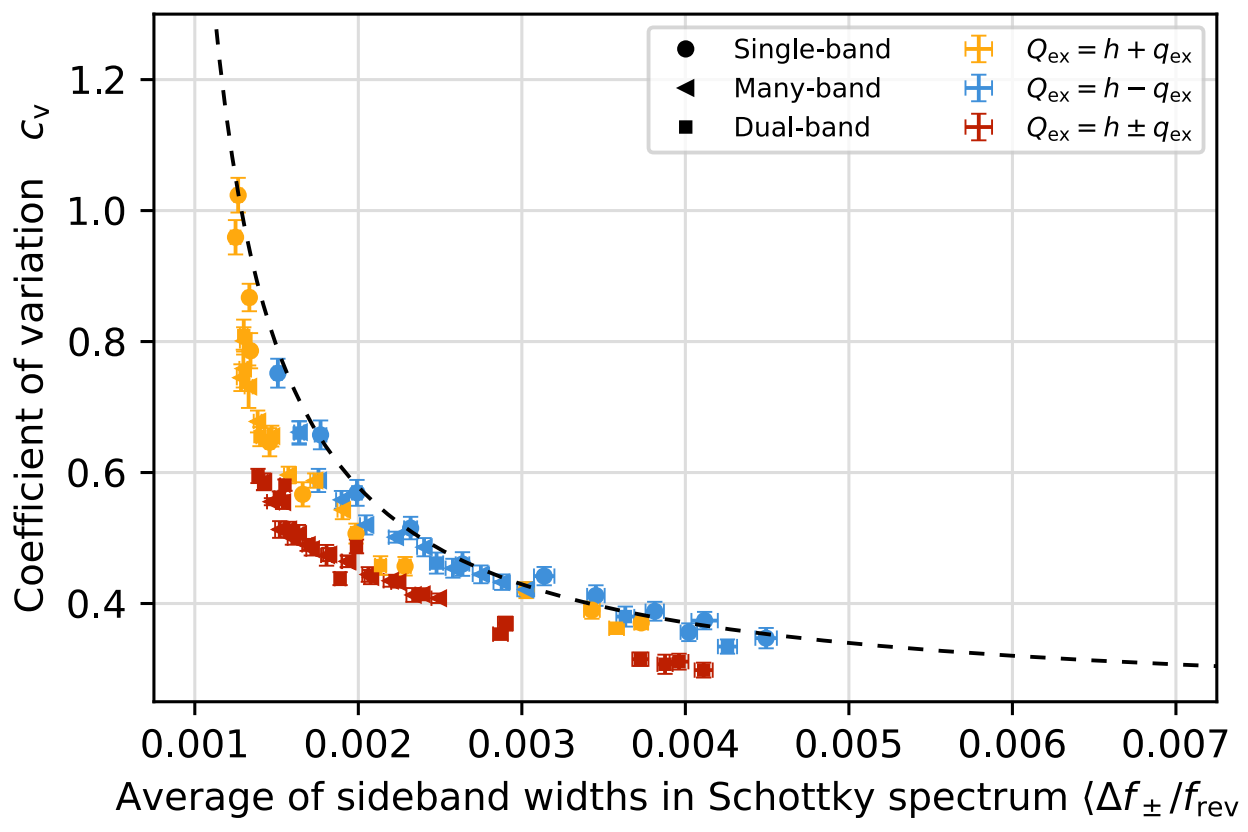

FIG. 9: Correlation between the average 1-sigma width of sidebands in the Schottky spectrum of a hollow beam (Fig. 5) and the spill quality resulting from their combined excitation (Fig. 8). The dashed line is the fit function Eq. (6) as obtained from Fig. 7.

In all cases, the combination of upper and lower sidebands results in an additional improvement of the spill quality compared to exciting either type alone (Fig. 9). This additional improvement goes beyond what is predicted by Eq. (6) using the average width of the involved sidebands, hinting that it is caused by a different mechanism. This can possibly be attributed to the fact that a detuning corresponds to a frequency change in opposite directions for upper and lower sidebands respectively. The simultaneous coupling of the excitation through these opposite bands leads to a more homogeneous excitation process which is less coherent and reduces the imprint of the excitation onto the spill beyond the average of the two single bands. This can be seen in Fig. 10 at the respective excitation band. The weaker imprint also leads to reduced spill fluctuations in the baseband and thus improves the spill quality.

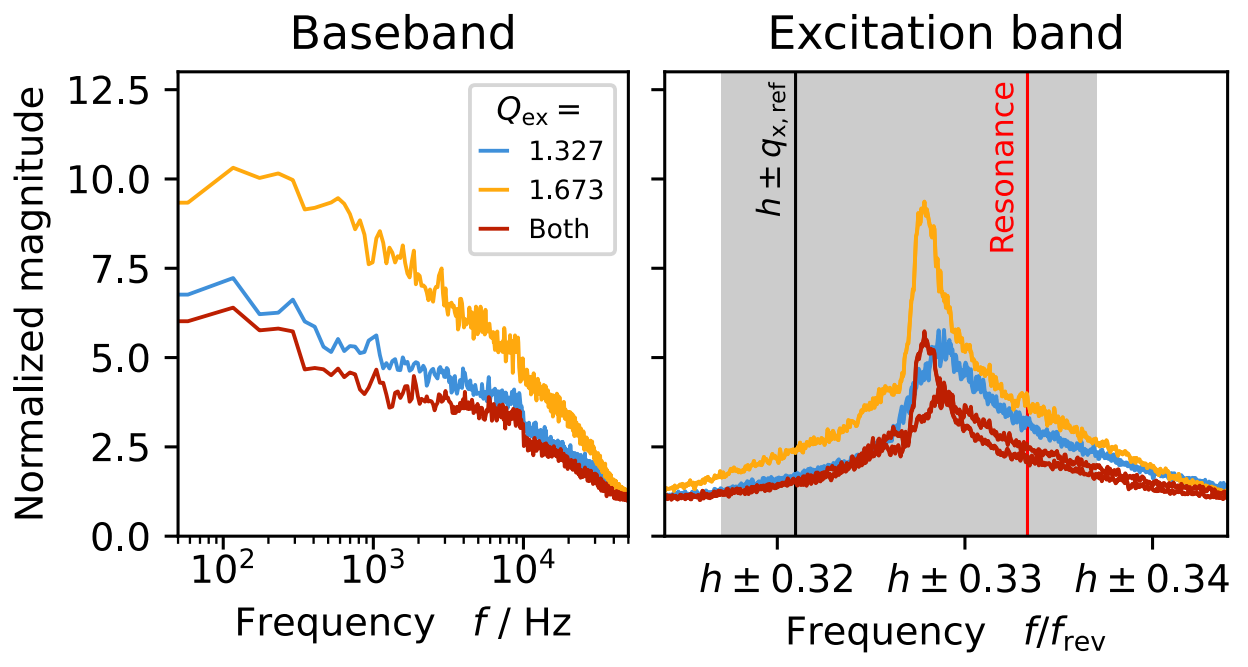

FIG. 10: Spill spectra at the baseband (left) and at the respective excitation band (right) for two single band excitations and the dual band case combining the two of them. Vertical lines mark the tune and resonance and the gray shading marks the extent of the excitation. Simulation result.

In conclusion, and subject to the parameters discussed in section II B, dual-band excitation at upper and lower sidebands of high harmonics is the most beneficial excitation method in terms of the spill quality. While slightly more excitation power is required, this method can simplify the design of excitation systems compared to the many-band case spanning many harmonics, since a much smaller system bandwidth is required. For example, exciting the upper and lower sideband of the 20th harmonic would require a system bandwidth of only 10 % of the tunable central frequency. The reduced demand in terms of frequency characteristics could allow for a higher impedance power coupling mechanisms with less power losses [17].

## V. EXPERIMENTAL INVESTIGATION

### A. Measurement setup

The spill quality and amount of pile-up were measured at the Heidelberg Ion Beam Therapy Center. Therefore, an ion optics with $Q_{x,ref} = 1.6802$ and $\xi_x = -0.655$ was used, and a $^{12}C^{6+}$ beam was extracted at $E_{kin} = 251\,\mathrm{MeV/u}$ ($f_{rev} = 2.8438\,\mathrm{MHz}$). While for treatment purposes the rf-KO extraction is performed with the beam bunched by the accelerating rf cavity, bunching was turned off for the measurement to simplify the beam dynamics interpretation and allow for a closer comparison to the simulations. The initial beam momentum spread was measured to be $\delta_{1\sigma} \approx 7 \cdot 10^{-5}$ using a longitudinal Schottky monitor. While the measurement conditions differ from the initial baseline simulations shown in Figs. 1, 2 and 3, they lie within the large parameter range considered in Fig. 4.

For the measurement, rf-KO extraction was performed with the RBPSK excitation and spill intensity feedback system used for treatment, and with a band-filtered noise excitation of manually adjusted amplitude. In both cases, a central frequency and bandwidth according to table I was used. A 500 W power amplifier with a bandwidth of 20 MHz was used to power the 50 Ω terminated stripline exciter. The exciter's electrode plates have a physical length of 0.75 m and an electrical length of 1.1 m each (measured between the vacuum feed-throughs). They are connected in series with cables and a transformer in between, which is designed to invert the signal polarity at frequencies below 2 MHz in order to create the transverse deflecting rf field between both electrodes. Given this serial connection scheme, an additional frequency dependent phase shift is introduced by the total length of the signal path between the plates. At about 17 MHz $\approx 6 f_{rev}$ the additional shift amounts to 180°, which including the transformer leads to a total phase shift of about 360° between the plates, effectively causing the transverse rf field to vanish. The field has another maximum at about 34 MHz $\approx 12 f_{rev}$, where the total phase shift amounts to 540°.

The spills were recorded with an Ionisation Chamber (IC) at 50 µs resolution for analysis of the spill quality. In addition, a BC400 plastic scintillator connected to a photomultiplier was installed downstream of the IC to record the last 18 ms of the spill with an oscilloscope at 0.32 ns resolution. The scintillator pulses were analyzed with individual particle pulse detection using the standard peak detection algorithm `signal.find_peaks` of the scipy library [29]. The resulting timestamps are used for the pile-up analysis.

### B. Measured spill quality and pile-up

The spill quality, measured at an average extraction rate of $5 \cdot 10^6$ particles/s with the machine settings as stated above is shown in Fig. 11 as a function of the central frequency of the single-band RBPSK excitation signal. The spill quality gradually improves for higher sidebands, and there is a small yet systematic difference between excitation at lower and upper betatron sidebands. There is a fair agreement of the measured spill quality with the corresponding simulations, considering that the number of particles in the simulation is reduced due to limited computational resources, which results in a shift of the values of $c_v$. Given the low beam momentum spread, the measured effect is small compared to the initial simulations with a larger momentum spread in Fig. 1. While the simulations assume an ideal frequency response of the excitation system, in the measurement it was not possible to extract the desired intensity at the $Q_{ex} = 6.327$ band due to the frequency characteristic of HIT's excitation system as explained above.

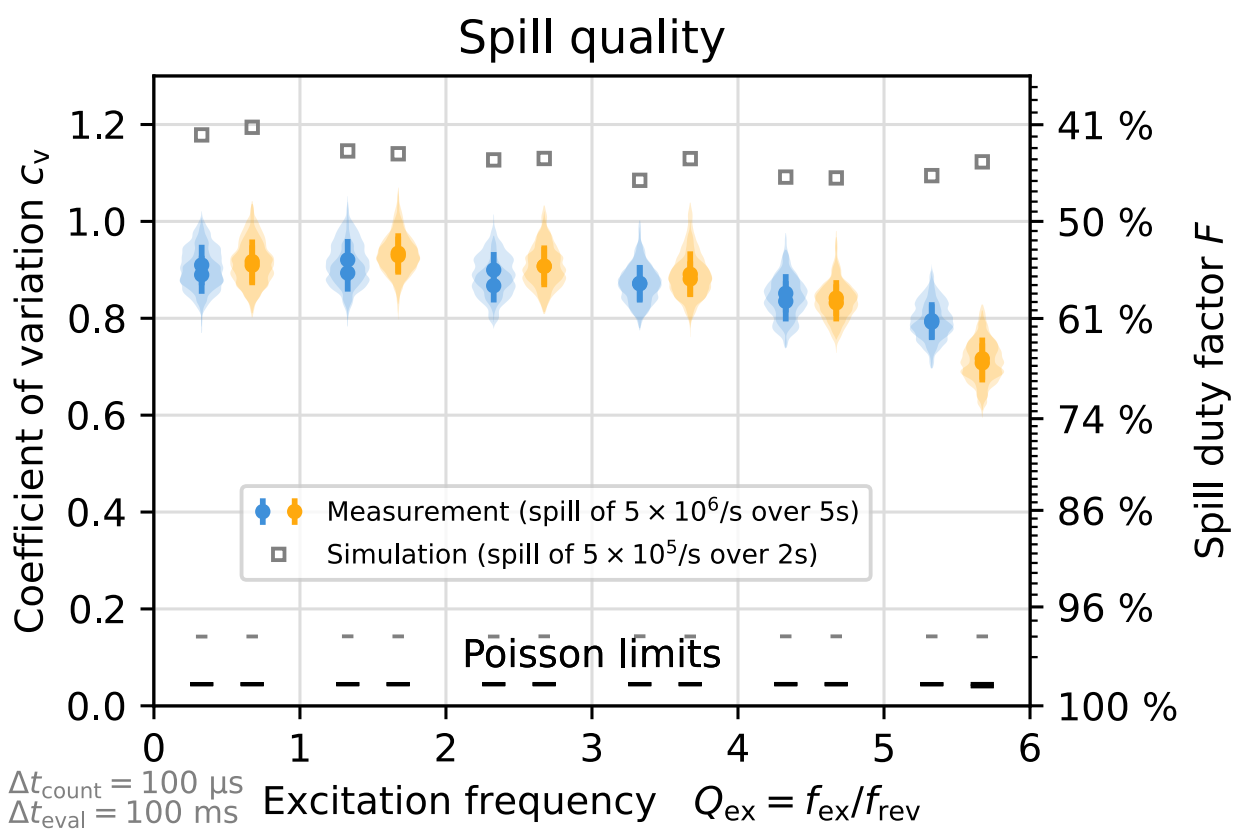

FIG. 11: Measured spill quality for single-band RBPSK excitation at different sidebands. For comparison, the simulation result as obtained for the same machine settings, but reduced spill rate of $5 \cdot 10^5$ 1/s and ideal frequency response of the exciter is shown.

For analysis of pile-up, individual particles were recorded with a scintillator under the same machine settings but at a lower extraction rate of $4 \cdot 10^6$ particles/s. The cumulative distribution of the measured time between consecutive particles is shown in Fig. 12 (left). In the figures, also a selected region of 2 µs of the raw signal with piled-up pulses is highlighted. It can be seen that the predominant periodicity of particles is defined by the excitation frequency, resulting in clustering at a lower excitation frequency. With the scintillator pulse full width at half maximum (FWMH) being about 25 ns, consecutive particles arriving within this period would pile-up. By evaluating the fraction of particles arriving within less than $\Delta t_{\text{pile-up}} = 20$ ns, the amount of pile-up is shown as a function of excitation frequency in Fig. 12 (right). As the excitation frequency increases with respect to the average particle rate, the clustering is avoided and pile-up reduces. Due to the small momentum spread in the experiment, the effect of spill quality improvement and associated reduction of excitation frequency imprint as discussed in section II and observed in Fig. 3 is weakened, such that the measured pile-up behaviour is dominated by the clustering frequency and no systematic difference between upper and lower sidebands is observed. This also suggests that the excitation frequency can be utilized as a free parameter to mitigate pile-up independent of the momentum spread: For large momentum spreads the mechanism detailed in sections II and III can be used to reduce pile-up by suppressing the imprint of the excitation onto the spill as shown in Fig. 3. For small momentum spreads the imprint can instead be exploited by matching the excitation frequency to the desired rate of the detector system in order to reduce pile-up by suppressing the clustering as shown in Fig. 12.

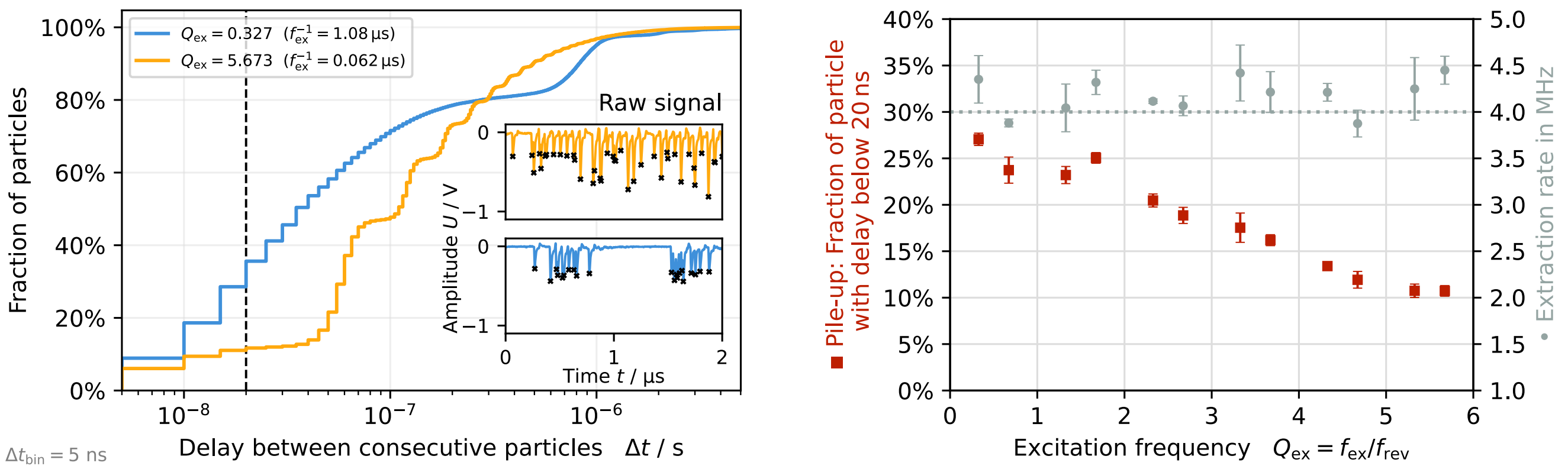

FIG. 12: Left: Cumulative distribution of particle arrival times for RBPSK excitation at $Q_{ex} = 0.327$ and 5.673. The insets show the raw detector signals with pulses corresponding to the detection of individual particles (marked with crosses). Right: Fraction of particles arriving within less than $\Delta t_{\text{pile-up}} = 20$ ns of one another (pile-up) as a function of the excitation frequency as well as the corresponding average extraction rate of about $4 \cdot 10^6$ particles/s.

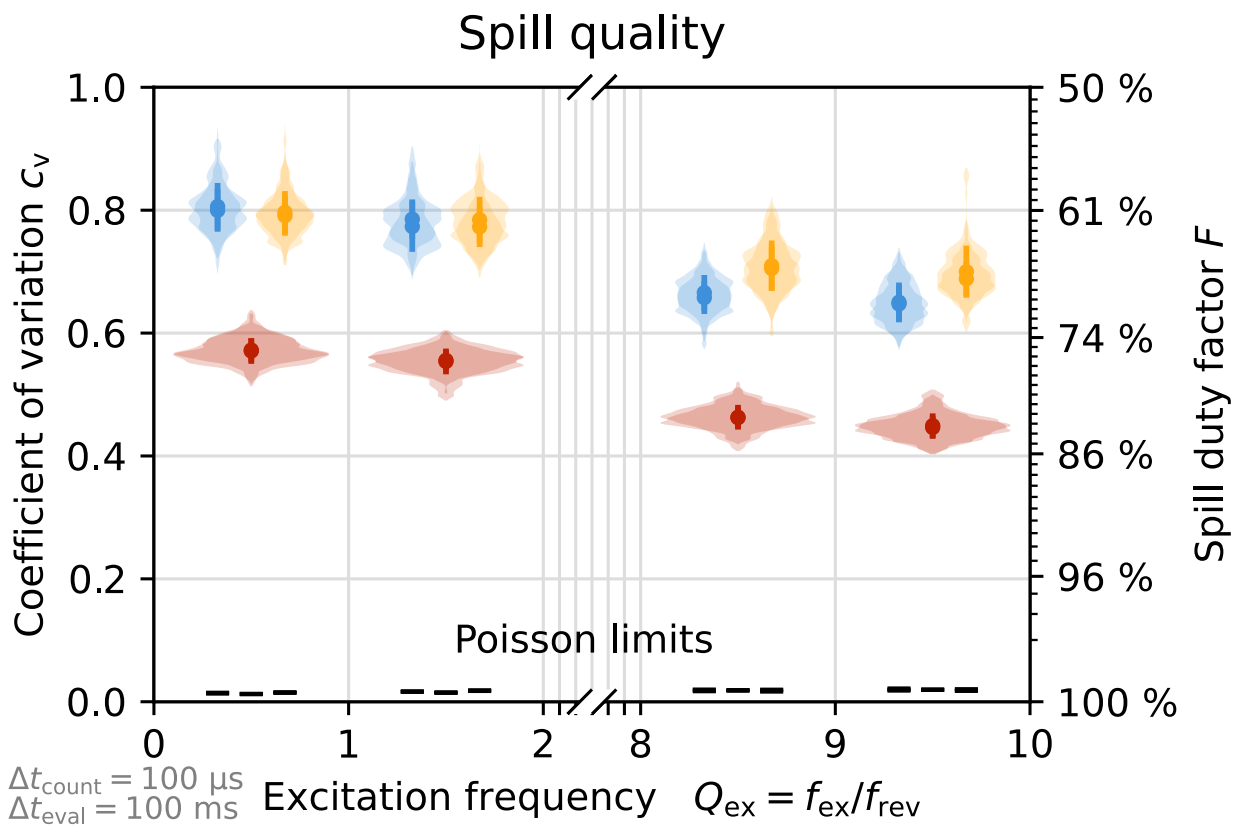

FIG. 13: Measurement of the spill quality for single-band noise excitation (blue and orange) and dual-band excitation (red) at sidebands of different harmonics. The dual-band cases are plotted in between the two neighbouring excitation bands they comprise.

Finally, a comparison of the measured spill quality of single- and dual-band excitation signals is shown in Fig. 13. In agreement with the simulations, an improvement is observed at sidebands of higher harmonics. Independent of the harmonic and despite the small momentum spread, a significant additional improvement is found by combining one upper and one lower sideband. As discussed above, the latter is the result of a weakened excitation imprint onto the spill.

## VI. CONCLUSION

Coasting beam extraction by exciting betatron sidebands at various harmonics individually as well as in different combinations was studied in simulations using the HIT synchrotron. A band-filtered noise signal with a bandwidth covering the frequency range from the respective zero-amplitude sideband to the nearby resonance is used, ensuring a good extraction efficiency. The simulations suggest a strong correlation of the spill quality with the spread of betatron frequencies of particles near the separatrix which are about to be extracted, observed as sideband width in the spectrum of hollow beams. The latter strongly depends on the beam and ion optical settings and is investigated for the chromaticity, momentum spread and slip factor. In particular, for beams with a sufficient momentum spread, the hollow-beam sideband width increases for sidebands at high harmonics, and depending on the chromaticity and slip factor can differ significantly for upper and lower betatron sidebands. By exciting the beam at such sidebands, perturbations of the rf-KO extraction process and the imprint of the excitation itself onto the spill are reduced, improving the spill quality and mitigating pile-up. For multi-band excitation, the average hollow-beam sideband width is determining for the spill quality, though due to the nonlinear nature of the resonant extraction the combined excitation of nearby upper and lower sidebands shows an additional improvement.

Beam experiments at HIT were performed to measure spill quality and pile-up. The observed trend of spill quality improvement with the betatron sideband is small, which — given the low momentum spread under the experimental conditions — is in line with the simulations, as the effect scales strongly with the momentum spread. As suggested by the simulations, the combined excitation of two sidebands shows an additional improvement of the measured spill quality. In the studied case, one upper and one lower sideband gives the best result. The experiment also shows a reduction of pile-up when exciting sidebands at higher harmonics, avoiding the clustered extraction of particles. Mitigation of pile-up is especially relevant for high rate physics experiments, for which choosing higher excitation frequencies is preferable.

The results suggest that excitation systems for rf-KO extraction should be designed with the possibility of applying higher harmonic excitation to the beam. This does not only include the generation and amplification of the rf signals, but also the geometry and driving schema of the exciter electrodes. For the commonly used stripline exciters, each electrode should be powered individually using identical signal path lengths to avoid the frequency dependent response observed in the HIT measurement. In addition, for non-relativistic beams, also the wavelength must be large compared to the electrode length, limiting the accessible frequencies to a typical order of 100 MHz. Exciting only a single or two sidebands at a higher harmonic can potentially simplify the design of rf amplifiers, since the required bandwidth is only a small fraction of the revolution frequency as opposed to currently used broadband amplifiers covering many revolution harmonics (as in [14,17]).

When high frequencies are not accessible due to limitations of existing excitation systems, spill improvement techniques utilizing multiple of the lower sidebands can be used. This includes the excitation of two or three of the lower sidebands (compare Fig. 13 and [18]), the usage of improved beam excitation signals [23] or the excitation of second order betatron sidebands [19]. While these techniques reduce low frequency fluctuations of the spill intensity, thus improving spill quality and reducing pile-up, a reduction of the excitation frequency imprint has not been reported so far. The same applies to bunched beam extraction, where the frequency and harmonics of the bunching cavity are imprinted onto the spill [30–32]. The present study shows that the undesired high frequency imprint can be mitigated when utilizing high frequency excitation at betatron sidebands of higher harmonics, due to the increased intrinsic width of these sidebands for particles near the separatrix. This has the benefit of a further spill quality improvement while allowing for an optimal choice of the excitation frequency to mitigate pile-up. In addition, simulations suggest that only a few dB of additional excitation power are required for the high frequency excitation, since the method does not rely on using narrowband or sinusoidal excitation, but instead uses efficient signals such as band-filtered noise covering the full sideband width.

## ACKNOWLEDGEMENTS

We thank E. C. Cortés García for valuable discussions on his studies at HIT and related simulations. The authors acknowledge the funding received from the European Union's Horizon 2020 Research and Innovation program under GA No. 101004730.

## APPENDIX: DERIVATION OF DETUNED SIDEBAND WIDTH

Starting from Eq. (4), the frequency contribution of a single particle with momentum offset $\delta$ and amplitude detuning $\mathcal{V}_\mathrm{a}$ is

$$\frac{f_\pm}{f_\mathrm{rev}} \approx \left[h \pm q_\mathrm{x} \pm \xi_\mathrm{x}\delta \pm \mathcal{V}_\mathrm{a}\left(1-\frac{\xi_\mathrm{x}\delta}{\Delta_\mathrm{res}}\right)\right](1+\eta\delta)$$

The partial derivatives at the expectation values $\langle\delta\rangle = 0$ and $\langle\mathcal{V}_\mathrm{a}\rangle \neq 0$ are

$$\begin{aligned}
\frac{\partial}{\partial\delta}\left(\frac{f_\pm}{f_\mathrm{rev}}\right) &\approx \left[h \pm q_\mathrm{x} \pm \langle\mathcal{V}_\mathrm{a}\rangle\right]\eta \pm \xi_\mathrm{x}\left(1-\frac{\langle\mathcal{V}_\mathrm{a}\rangle}{\Delta_\mathrm{res}}\right) \pm 2\xi_\mathrm{x}\eta\left[1-\frac{\langle\mathcal{V}_\mathrm{a}\rangle}{\Delta_\mathrm{res}}\right]\langle\delta\rangle \\
&\approx \left[h \pm q_\mathrm{x} \pm \langle\mathcal{V}_\mathrm{a}\rangle\right]\eta \pm \xi_\mathrm{x}\left(1-\frac{\langle\mathcal{V}_\mathrm{a}\rangle}{\Delta_\mathrm{res}}\right)
\end{aligned}$$

$$\frac{\partial}{\partial\mathcal{V}_\mathrm{a}}\left(\frac{f_\pm}{f_\mathrm{rev}}\right) \approx \pm\left(1-\frac{\xi_\mathrm{x}\langle\delta\rangle}{\Delta_\mathrm{res}}\right)\left(1+\eta\langle\delta\rangle\right) \approx \pm 1$$

$$\frac{\partial^2}{\partial\delta\,\partial\mathcal{V}_\mathrm{a}}\left(\frac{f_\pm}{f_\mathrm{rev}}\right) \approx \pm\left(\eta-\frac{\xi_\mathrm{x}}{\Delta_\mathrm{res}}\right) \mp 2\frac{\xi_\mathrm{x}\eta}{\Delta_\mathrm{res}}\langle\delta\rangle \approx \pm\left(\eta-\frac{\xi_\mathrm{x}}{\Delta_\mathrm{res}}\right)$$

Treating $\delta$ and $\mathcal{V}_\mathrm{a}$ as independent random variables with standard deviations of $\delta_{1\sigma}$ and $\Delta\mathcal{V}_\mathrm{a}$ respectively and collecting only the linear contributions of $\delta$ and $\mathcal{V}_\mathrm{a}$ as well as the bilinear contribution of the product $\delta\mathcal{V}_\mathrm{a}$, the usual squared sum term follows as

$$\frac{\Delta f_\pm}{f_\mathrm{rev}} \approx \sqrt{\left[\frac{\partial}{\partial\delta}\left(\frac{f_\pm}{f_\mathrm{rev}}\right)\delta_{1\sigma}\right]^2 + \left[\frac{\partial}{\partial\mathcal{V}_\mathrm{a}}\left(\frac{f_\pm}{f_\mathrm{rev}}\right)\Delta\mathcal{V}_\mathrm{a}\right]^2 + \left[\frac{\partial^2}{\partial\delta\,\partial\mathcal{V}_\mathrm{a}}\left(\frac{f_\pm}{f_\mathrm{rev}}\right)\delta_{1\sigma}\Delta\mathcal{V}_\mathrm{a}\right]^2}$$

Adding the empirical additive broadening term $\Delta\mathcal{V}_\mathrm{p}$ to account for the phase detuning as motivated in the text and adding an empirical correction factor to the bilinear term yields Eq. (5):

$$\frac{\Delta f_\pm}{f_\mathrm{rev}} \approx \Delta\mathcal{V}_\mathrm{p} + \sqrt{\left[\left(h \pm q_\mathrm{x} \pm \langle\mathcal{V}_\mathrm{a}\rangle\right)\eta \pm \xi_\mathrm{x}\left(1-\frac{\langle\mathcal{V}_\mathrm{a}\rangle}{\Delta_\mathrm{res}}\right)\right]^2\delta_{1\sigma}^2 + \Delta\mathcal{V}_\mathrm{a}^2 + c_\mathrm{cor}\left(\eta-\frac{\xi_\mathrm{x}}{\Delta_\mathrm{res}}\right)^2\delta_{1\sigma}^2\Delta\mathcal{V}_\mathrm{a}^2}$$

The bilinear correction factor $c_\mathrm{cor}$ is a fit parameter which improves the matching for different values of the chromaticity by correcting for the approximation of the detuning distribution by a mean and standard deviation.

---

[1] K. Hiramoto and M. Nishi, Resonant Beam Extraction Scheme with Constant Separatrix, Nucl. Instrum. Methods Phys. Res., Sect. A **322**, 154 (1992).

[2] M. Tomizawa et al., Slow Beam Extraction at TARN II, Nucl. Instrum. Methods Phys. Res., Sect. A **326**, 399 (1993).

[3] M. Benedikt, P. J. Bryant, L. Badano, M. Crescenti, P. Holy, A. T. Maier, M. Pullia, S. Rossi, and P. Knaus, Proton-Ion Medical Machine Study (PIMMS): Part I, technical report No. CERN/PS99–10(DI), 1999, https://cds.cern.ch/record/385378.

[4] C. Belver-Aguilar, A. Faus-Golfe, F. Toral, and M. J. Barnes, Stripline Design for the Extraction Kicker of Compact Linear Collider Damping Rings, Phys. Rev. ST Accel. Beams **17**, 71003 (2014).

[5] M.-R. Mohammadian-Behbahani and S. Saramad, A Comparison Study of the Pile-up Correction Algorithms, Nucl. Instrum. Methods Phys. Res., Sect. A **951**, 163013 (2020).

[6] R. Singh, P. Forck, P. Boutachkov, S. Sorge, and H. Welker, Slow Extraction Spill Characterization From Micro to Milli-Second Scale, J. Phys. Conf. Ser. **1067**, 72002 (2018).

[7] S. van der Meer, Stochastic Extraction, A Low-Ripple Version of Resonant Extraction, technical report No. CERN-PS-AA-78-6, 1978, https://cds.cern.ch/record/2830887.

[8] R. Cappi and C. Steinbach, Low Frequency Duty Factor Improvement for the CERN PS Slow Extraction Using RF Phase Displacement Techniques, IEEE Trans. Nucl. Sci. **28**, 2806 (1981).

[9] G. Molinari and H. Mulder, *The Improved Ultra-Slow Extraction Noise System at LEAR*, in *Proc. 4th European Particle Accelerator Conf.* (London, UK, 1994), https://cds.cern.ch/record/265947.

[10] K. Noda, T. Furukawa, S. Shibuya, T. Uesugi, M. Muramatsu, M. Kanazawa, E. Takada, and S. Yamada, Advanced RF-KO Slow-Extraction Method for the Reduction of Spill Ripple, Nucl. Instrum. Methods Phys. Res., Sect. A **492**, 253 (2002).

[11] K. Noda, T. Furukawa, S. Shibuya, M. Muramatsu, T. Uesugi, M. Kanazawa, M. Torikoshi, E. Takada, and S. Yamada, Source of Spill Ripple in the RF-KO Slow-Extraction Method with FM and AM, Nucl. Instrum. Methods Phys. Res., Sect. A **492**, 241 (2002).

[12] T. Furukawa and K. Noda, Contribution of Synchrotron Oscillation to Spill Ripple in RF-Knockout Slow-Extraction, Nucl. Instrum. Methods Phys. Res., Sect. A **539**, 44 (2005).

[13] T. Nakanishi and K. Tsuruha, Simulation Study of Beam Extraction from a Synchrotron Using Colored Noise with Digital Filter, Nucl. Instrum. Methods Phys. Res., Sect. A **608**, 37 (2009).

[14] T. Nakanishi, Dependence of a Frequency Bandwidth on a Spill Structure in the RF-Knockout Extraction, Nucl. Instrum. Methods Phys. Res., Sect. A **621**, 62 (2010).

[15] K. Mizushima, T. Shirai, T. Furukawa, and K. Noda, Making Beam Spill Less Sensitive to Power Supply Ripple in Resonant Slow Extraction, Nucl. Instrum. Methods Phys. Res., Sect. A **638**, 19 (2011).

[16] R. Singh, P. Forck, and S. Sorge, Reducing Fluctuations in Slow-Extraction Beam Spill Using Transit-Time-Dependent Tune Modulation, Phys. Rev. Appl. **13**, (2020).

[17] T. Yamaguchi, Y. Okugawa, T. Shiokawa, T. Kurita, and T. Nakanishi, Slow Beam Extraction from a Synchrotron Using a Radio Frequency Knockout System with a Broadband Colored Noise Signal, Nucl. Instrum. Methods Phys. Res., Sect. B **462**, 177 (2020).

[18] E. C. Cortés García, E. Feldmeier, M. Galonska, C. Schömers, M. Hun, S. Brons, R. Cee, S. Scheloske, A. Peters, and T. Haberer, Optimization of the Spill Quality for the Hadron Therapy at the Heidelberg Ion-Beam Therapy Centre, Nucl. Instrum. Methods Phys. Res., Sect. A **1040**, 167137 (2022).

[19] P. Niedermayer and R. Singh, Excitation of Nonlinear Second Order Betatron Sidebands for Knock-out Slow Extraction at the Third-Integer Resonance, Phys. Rev. Accel. Beams **27**, 82801 (2024).

[20] G. Iadarola et al., *Xsuite: An Integrated Beam Physics Simulation Framework*, in *Proc. 68th Adv. Beam Dyn. Workshop High-Intensity High-Brightness Hadron Beams* (JACoW Publishing, Geneva, Switzerland, 2023), pp. 73–80, 10.18429/JACoW-HB2023-TUA2I1.

[21] C. Kleffner, D. Ondreka, U. Weinrich, F. D. McDaniel, and B. L. Doyle, *The Heidelberg Ion Therapy (HIT) Accelerator Coming into Operation*, in *AIP Conf. Proc.*, Vol. 1099 (AIP Conference Proceedings, Fort Worth, TX, USA, 2009), pp. 426–428, 10.1063/1.3120065.

[22] S. White, E. Maclean, and R. Tomás, Direct Amplitude Detuning Measurement with AC Dipole, Phys. Rev. ST Accel. Beams **16**, (2013).

[23] P. Niedermayer and R. Singh, Excitation Signal Optimization for Minimizing Fluctuations in Knock out Slow Extraction, Sci Rep **14**, 10310 (2024).

[24] C. Krantz et al., *Slow Extraction Techniques at the Marburg Ion-Beam Therapy Centre*, in *Proc. 9th Int. Particle Accelerator Conf.* (JACoW Publishing, Vancouver, BC, Canada, 2018), pp. 1084–1086, 10.18429/JACOW-IPAC2018-TUPAL036.

[25] R. Singh, P. Boutachkov, P. Forck, P. Kowina, P. Schmid, A. Stafiniak, and H. Welker, Study of SIS-18 Spill Structure by Introducing External Ripples, GSI Scientific Report 2016 447 (2017).

[26] D. Boussard, *Schottky Noise and Beam Transfer Function Diagnostics*, in *Proc. CERN Accelerator School: Advanced Accelerator Physics Course* (CERN, Rhodes, Greece, 1993, 1995), pp. 749–782, 10.5170/CERN-1995-006.749.

[27] E. C. Cortés García, P. Niedermayer, R. Singh, R. Taylor, E. Benedetto, E. Feldmeier, M. Hun, and T. Haberer, Interpretation of the Horizontal Beam Response near the Third Integer Resonance, Phys. Rev. Accel. Beams **27**, 124001 (2024).

[28] P. Niedermayer, Transverse Excitation for Beam Diagnostics and Slow Extraction from Synchrotrons, Doctoral dissertation, 2025, 10.21248/gups.91300.

[29] P. Virtanen et al., SciPy 1.0: Fundamental Algorithms for Scientific Computing in Python, Nat. Methods **17**, 261 (2020).

[30] K. Gross, B. Zipfel, D. Lens, H. Klingbeil, J. Schmidt, P. Hülsmann, P. Spiller, R. Balss, T. Winnefeld, and U. Laier, *Development of a Spill-Structure Manipulation Cavity and First Experiment with Beam in SIS18*, in *Proc. 15th Int. Particle Accelerator Conf.* (JACoW Publishing, Nashville, TN, USA, 2024), pp. 1432–1435, 10.18429/JACOW-IPAC2024-TUPR07.

[31] S. Sorge, P. Forck, and R. Singh, Spill Ripple Mitigation by Bunched Beam Extraction with High Frequency Synchrotron Motion, Phys. Rev. Accel. Beams **26**, 14402 (2023).

[32] J. Yang, *Nano-Second Time Scale Measurements of the Particle Arrival Times*, Slow Extraction Workshop, MedAustron, Wiener Neustadt, (unpublished).